\documentclass[%
 aip,
 amsmath,amssymb,
 reprint,%
]{revtex4-1}
\usepackage{comment}
\usepackage{graphicx}
\usepackage{dcolumn}
\usepackage{bm}
\usepackage{hyperref}
\usepackage{xcolor}
\hypersetup{
    colorlinks=true,
    linkcolor=blue,
    citecolor=blue,
    urlcolor=cyan,
}
\usepackage{graphicx}
\usepackage{dcolumn}
\usepackage{bm}
\usepackage[utf8]{inputenc}
\usepackage[T1]{fontenc}
\usepackage{mathptmx}

\begin{document}

\preprint{APS/123-QED}

\title{Sustained coherent spin wave emission using frequency combs}

\author{A. A. Awad}
\thanks{These authors contributed equally to this work.}%
\affiliation{Physics Department, University of Gothenburg, 412 96 Gothenburg, Sweden.}
\author{S. Muralidhar}
\thanks{These authors contributed equally to this work.}%
\affiliation{Physics Department, University of Gothenburg, 412 96 Gothenburg, Sweden.}
\author{A. Alem\'an}
\thanks{These authors contributed equally to this work.}%
\affiliation{Physics Department, University of Gothenburg, 412 96 Gothenburg, Sweden.}
\author{R. Khymyn}
\affiliation{Physics Department, University of Gothenburg, 412 96 Gothenburg, Sweden.}
\author{M. Dvornik}
\affiliation{Physics Department, University of Gothenburg, 412 96 Gothenburg, Sweden.}
\author{D. Hanstorp}
\affiliation{Physics Department, University of Gothenburg, 412 96 Gothenburg, Sweden.}
\author{J. \AA kerman}
\email{johan.akerman@physics.gu.se}
\affiliation{Physics Department, University of Gothenburg, 412 96 Gothenburg, Sweden.}
\affiliation{Materials and Nano Physics, School of Engineering Sciences, KTH Royal Institute of Technology, Electrum 229, 164 40 Kista, Sweden.}

\begin{abstract}
We demonstrate sustained coherent emission of spin waves in NiFe films using rapid demagnetization from high repetition rate femtosecond laser pulse trains. As the pulse separation is shorter than the magnon decay time, magnons having a frequency equal to a multiple of the 1 GHz repetition-rate are coherently amplified. Using scanning micro-Brillouin Light Scattering (BLS) we observe this coherent amplification as strong peaks spaced 1 GHz apart. The BLS counts vs.~laser power exhibit a stronger than parabolic dependence consistent with counts being proportional to the square of the magnetodynamic amplitude, and the demagnetization pulse strength being described by a Bloch law. Spatial spin wave mapping demonstrates how both localized and propagating spin waves can be excited, and how the propagation direction can be directly controlled. Our results demonstrate the versatility of BLS spectroscopy for rapid demagnetization studies and enable a new platform for photo-magnonics where sustained coherent spin waves can be utilized.
\end{abstract}

\maketitle

\section{Introduction}
Magnonics has emerged as a central research topic in nanomagnetism, with rich physics and an increasing number of novel phenomena thanks to the unique field-tunable properties of spin waves (SWs) and a wide range of metallic and insulating magnetic materials.\cite{Kruglyak2010,Magnonics2013, Neusser2009, Chumak2015}
As the wavelength of SWs can be several orders of magnitude smaller than its electromagnetic radiation counterparts at the same frequency, the possibility of scaling down high-frequency devices using magnonics offers excellent prospects for miniaturization.\cite{Neusser2009, Yu2016}

SWs can be excited using a wide range of mechanisms and techniques. While the most straightforward and conventional SW generation mechanism is that of an externally applied microwave field using RF antennas,\cite{Wolf1961,Ganguly1975,DemidovAPL2008} the more recent spin transfer-torque and spin Hall effects generated by direct currents through nanodevices have made it possible to generate truly short wavelength, highly non-linear, and very high intensity SWs on the nanoscale.\cite{demidov2010ntm, BonettiPRL2010, madami2011ntn, chen2016ieee, Houshang2015ntn, Philipp_2017, Houshang2018ntcomm, Divinskiy2018, Fulara2019} SWs can also be generated optically using focused femtosecond laser (fs-laser) pulses inducing rapid demagnetization\cite{Beaurepaire1997, Koopmans_nmat_2010, Rudolf2012,Lambert2014science, Mangin2014} of the local magnetization.\cite{JU1999PRL,VanKampen2002,Zhang2002PRL,Vomir2005PRL,Kruglyak2005PRB,Kimel2006PRB,Lambert2014science} 

Single-pulse excitation schemes, where the system relaxes into equilibrium before the arrival of the second pulse, have been studied extensively in metals\cite{VanKampen2002,Au2013,Yun2015,Iihama2016b,Kamimaki2017} and dielectrics\cite{Kimel2005,Satoh2012,Deb2015}. SWs excited from an individual pulse will have damped out well before the arrival of the next pulse since  the pump pulses are usually separated by about 12~ns or more, corresponding to a repetition rate of 80~MHz or lower, and the SW decay time in typical ferromagnetic metals is a few nanoseconds. For any chance of generating continuous spin waves using optical means hence requires much shorter times in between pulses in order to overcome the damping.
\begin{figure*}
    \begin{center}
        \includegraphics[width=0.91\linewidth]{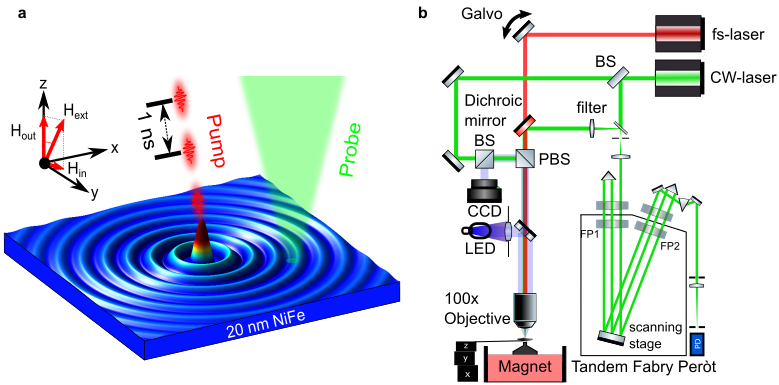}
        \caption{\textbf{Schematic of the experiment and the experimental setup.} \textbf{a.} The sample is pumped with a red (816 nm) 120 fs, 1 GHz-pulsed laser and probed by a continuous green laser (532 nm); the relative distance between the pump and the probe can be scanned over $\pm$40 $\mu$m. The magnetic field $H_{ext}$ is applied at an oblique out-of-plane angle of 82$^\circ$. 
        \textbf{b.} The optical setup of the pump-probe experiment: BS -- 50:50 Beam Splitter, PBS -- Polarizing Beam Splitter. The sample is placed right below a 100x objective with NA = 0.75 to achieve diffraction limited focusing. The back-scattered beam is analysed using a 6-pass Tandem Fabry-Pérot Interferometer (TFPI) and detected with a single channel avalanche photodiode (APD).}
        \label{fig:1}
    \end{center}
\end{figure*}

The first steps towards reducing the time between consecutive pulses were taken by employing pump-probe techniques with dual pump pulses. By tailoring the time delay between two such pump pulses, the precession in both single-\cite{Bossini2016,Hansteen2006} and multi-layer\cite{BERK2019IEEE} systems could either be quenched or amplified. While carefully controlling the phase relation allowed for detailed selectivity of which SWs to excite and further amplify, the overall duty cycle was however not improved.

A much more accessible approach is instead to increase the fs-laser repetition rate to approach the time scale of the SW decay. As frequency comb based fs-lasers with GHz repetition rates have recently become commercially available, the very first studies of high repetition rate SW excitations in thick extended Yttrium Iron Garnet (YIG) films\cite{Jackl2017,Savochkin2017} were reported. Using the inverse Faraday effect for excitation with a 10 $\mu$m laser spot size, and a conventional Time Resolved Magneto-Optical Kerr effect (TR-MOKE) pump-probe technique for detection, the authors were able to demonstrate coherently amplified excitation of long wavelength SWs whose phase relation matched the time between consecutive pulses. 

These pioneering works raise a number of intriguing questions. First, can high repetition rate fs-lasers also be efficiently used to coherently amplify SWs in metallic thin films and devices where the SW damping is higher? Second, can the high-repetition rate lasers enable the use of time-averaged techniques such as frequency resolved BLS microscopy in the study of ultrafast magnetization dynamics? Finally, if the pump laser and the BLS laser are focused down to their diffraction limits, can both localized and propagating SWs with both small and large wave vectors be excited, detected, and controlled? 

Here we attempt to answer these questions. Using a unique Brillouin light scattering (BLS) microscope, where we combine a diffraction limited BLS SW detection scheme with a diffraction limited high repetition rate (1 GHz) fs-laser inducing rapid demagnetization, we demonstrate continuous coherent SW emission over a wide range of fields and frequencies. The high sensitivity of our BLS microscope resolves both localized and propagating SWs. By varying the separation between the pump and probe laser spots, the spatial profiles and directions of the SWs can be elucidated in detail. In contrast to TR-MOKE measurements, where the signal increases linearly with laser power, we observe a nonlinear, stronger than square dependence of the BLS counts. The coherent excitation of magnons is further corroborated by the observation that this highly efficient SW generation only occurs at higher harmonics of the 1 GHz laser repetition rate. In other words, only SWs that are \emph{in phase} with the incoming laser pulses are coherently amplified, while all other SWs are left unaffected. The deviations from a square dependence can be fully accounted for by incorporating a Bloch law dependence of how the step in demagnetization field relates to the instantaneous magnon temperature.\cite{Atxitia2010,Mendil2014} By tuning the SW dispersion, through the magnitude and direction of the applied magnetic field, we can furthermore choose a specific wave vector to be coherently amplified, and also steer the direction of the amplified SW propagation. Our results clearly demonstrate the versatility and benefits of using spatially resolved BLS microscopy for the study of rapid demagnetization.

\section{Experimental details}

\subsection{Sample fabrication}
Permalloy (Ni$_{80}$Fe$_{20}$) thin films  were sputtered onto clean sapphire substrates ($c$-plane) by dc magnetron sputtering under a 3~mTorr Ar atmosphere at ultra-high vacuum with a $1.5\times 10^{-8}$~mTorr base pressure. Sapphire was chosen for its known negligible absorbance at the excitation laser wavelength and high thermal conductivity. The films were capped in-situ with 2~nm SiO$_2$ to prevent the magnetic layer from oxidation.  

\subsection{Frequency comb excitation and BLS microscope}
Our pump-probe experiment is shown schematically in Fig.~\ref{fig:1}.a: The pump beam was produced by a commercial Ti:Sa mode-locked laser with a 1~GHz repetition rate at a wavelength of 816~nm,  a 30~fs Fourier-limited pulse duration and pulse energies up to 1nJ. The laser pulse stretches during propagation in the optical system (Fig~\ref{fig:1}.b) and reaches the sample with a duration of 120~fs.  The laser was focused close to the diffraction limit using a 100x microscope objective with a NA of 0.75. 
The magneto-dynamics was probed using BLS microscopy.
\cite{Sebastian2015} A single-frequency continuous wave laser at 532 nm was focused to a near diffraction limited spot on the sample through the same objective as used for the pump beam.  The probe beam undergoes a frequency shift due to scattering from spin waves.  Back scattered light was collected and filtered using a polarizer and analyzed in a 6-pass Tandem Fabry-Perot Interferometer (TFPI) and detected using a single channel avalanche photodiode. The scattered light carries the phase, frequency and wavevector information from the spin wave. The spatial resolution is obtained at the expense of wavevector resolution by using the high-resolution microscope objective which forms a tight focus of the probe beam. The optical system is equipped with a pair of galvo-mirrors and lenses that allow the pump beam to be scanned over the sample to change the lateral distance between the pump and the probe beams.

\section{Results and Discussion}
\subsection{Field and power-dependent spectrum:}
\begin{figure}[h]
    \begin{center}
        \includegraphics[width=\linewidth]{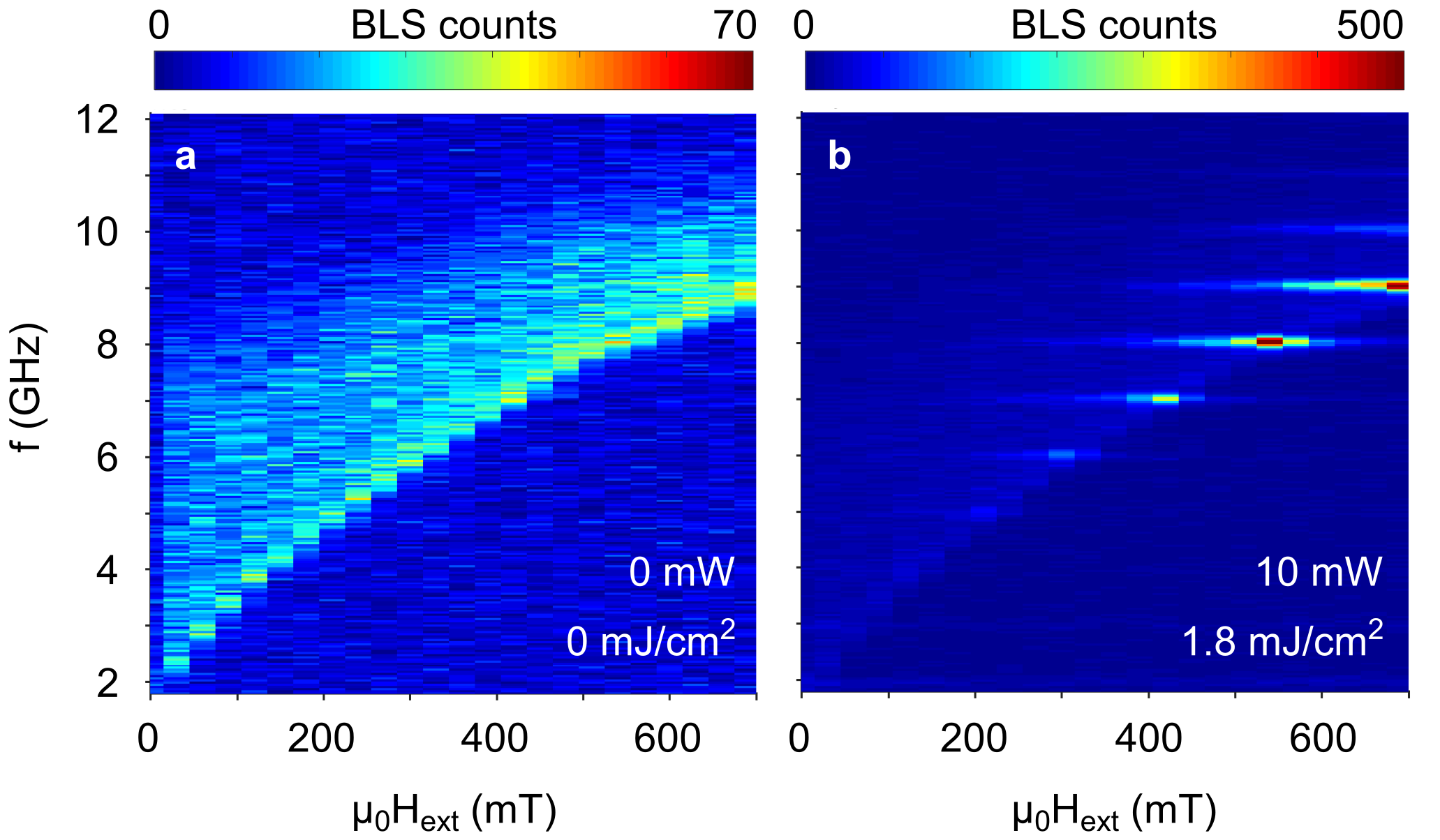}
        \caption{\textbf{Thermal and laser driven spin wave excitation in a 20 nm NiFe thin film.} 
        \textbf{a.} Thermal SW spectrum vs.~magnetic field applied at a 82$^\circ$ out-of-plane angle. \textbf{b.} SW spectrum when the same film is irradiated with the fs-laser pulse train at a 1.8 mJ/cm$^2$ fluence.}
        \label{fig:BLsThermalVsLaser}
    \end{center}
\end{figure}
Fig.~\ref{fig:BLsThermalVsLaser}.a shows the typical field-dependence of the thermal SW spectrum for the Py thin film sample when the pump laser is off. The sharp lower cut-off of the SW band can be clearly seen as it follows a Kittel-like frequency-vs-field dependence. A more gradual decay of the BLS counts is observed at higher SW frequencies as the SW wave-vector approaches the resolution limit of the BLS. Fig.~\ref{fig:BLsThermalVsLaser}.b shows the corresponding spectrum at the same location when the fs-laser pulse train is turned on with an average power of 10~mW, (pump fluence of 1.8~mJ/cm$^2$). The spectrum changes character entirely and instead of being largely featureless, a number of distinct peaks now appear at the harmonics of the 1 GHz repetition rate. The strongest peaks have about an order of magnitude higher intensity than the thermal background. While the peaks seem to fall close to the bottom of the original SW band, there is now also sizable BLS counts well below the gap at these harmonics.

\begin{figure}[t]
    \begin{center}
        \includegraphics[width=0.96\linewidth]{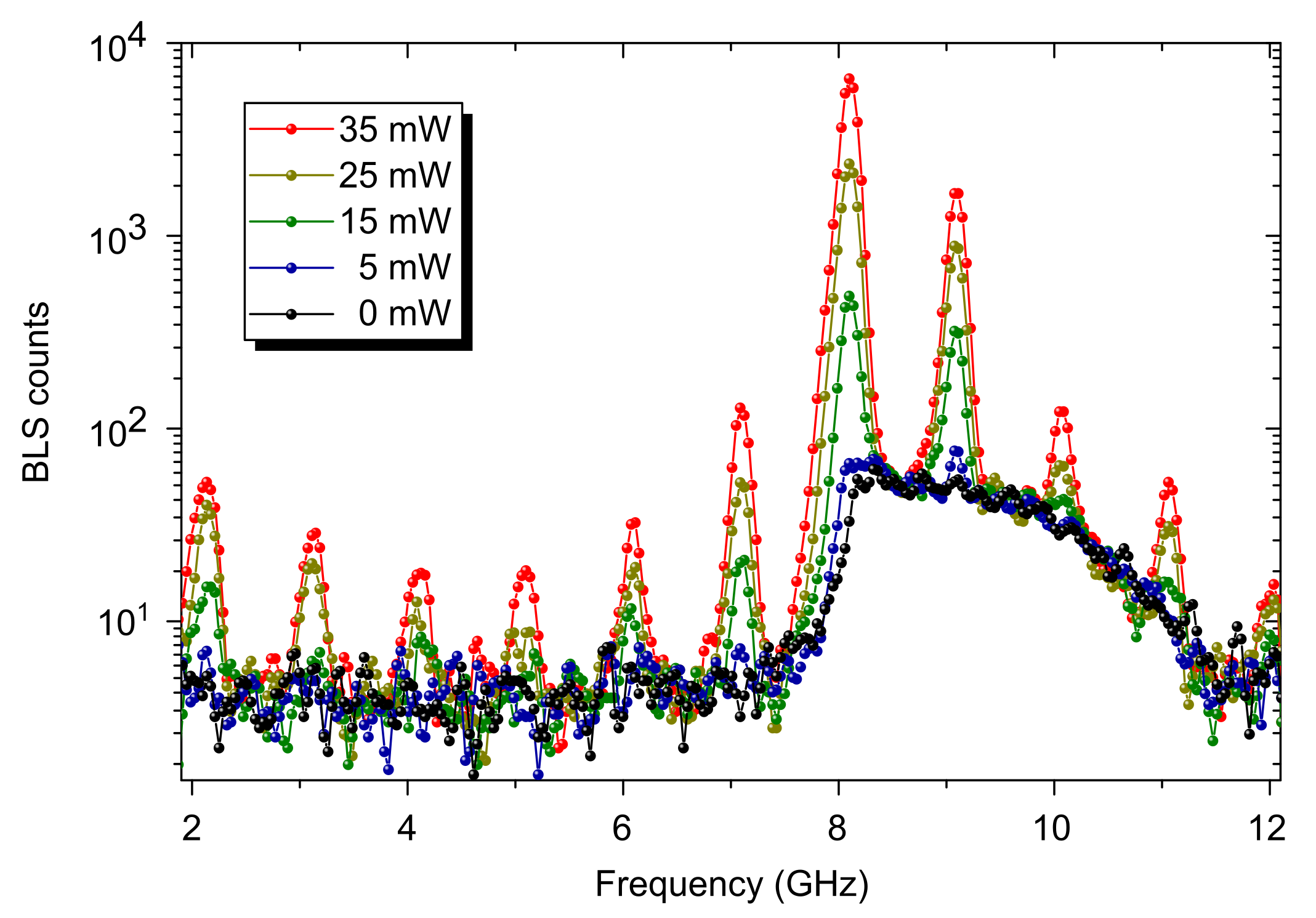}
        \caption{\textbf{Spin wave spectrum at different laser powers.} 
        BLS counts vs.~frequency in a field of 600~mT for only thermal SWs (black) and at four different pump powers showing pronounced peaks at the harmonics of the 1 GHz repetition rate.
        }
        \label{fig:BLSCountsVsF}
    \end{center}
\end{figure}

To better discern all features of the laser induced SW intensity, we show in Fig.~\ref{fig:BLSCountsVsF} a detailed plot of the BLS counts vs.~frequency at a field magnitude of 600 mT for four different laser powers together with the thermal SW background. We first note that 5 mW seems to be the approximate threshold for any noticeable additional SW intensity compared to the thermal SWs. At this lowest power level, we can observe additional BLS counts at 7, 8, and 9 GHz, but not at any other harmonics. At the higher power levels, pronounced SW peaks are observed at all harmonics, although the signal at these three frequencies remain the most strongly affected.

A number of qualitatively different frequency regions can be observed based on the data taken at high laser power. At low frequency we find that the maximum BLS counts decrease exponentially with harmonic number 2--4. At intermediate harmonic numbers (5--8) the BLS counts increase rapidly and reach a strong maximum at the FMR frequency of about 8 GHz. For yet higher harmonics, the counts again fall, most precipitously between the 9\textsuperscript{th} and the 10\textsuperscript{th} harmonic, above which the impact of the fs-laser remains limited compared to the thermal SWs.

\begin{figure}[b]
    \begin{center}
        \includegraphics[width=0.9\linewidth]{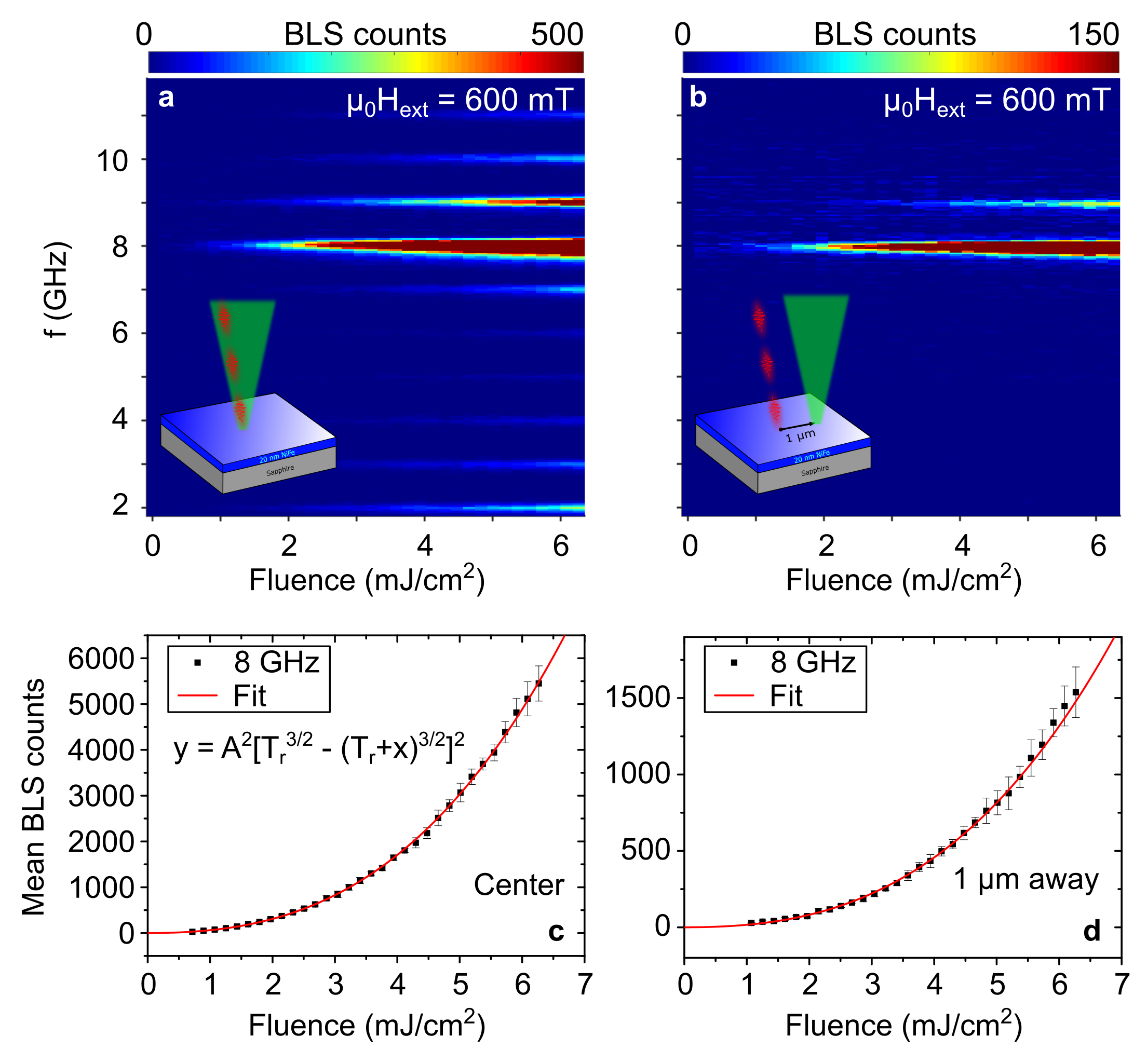}
        \caption{\textbf{Laser driven local and propagating spin waves.} 
        \textbf{a.} Power dependence of the BLS counts at the pump spot. \textbf{b.} Power dependence of the BLS counts when the probe is 1 $\mu$m away from the pump, showing SW propagation at 8 and 9~GHz. \textbf{c.} Peak counts at 8 GHz vs.~fluence at the pump spot together with a fit to our model described in the text. \textbf{d.} Peak counts at 8 GHz vs.~fluence 1 $\mu$m away from the pump spot together with a fit.}
        \label{fig:LocalVsPropSWs}
    \end{center}
\end{figure}

To complement the data shown for overlapping pump and probe lasers in Fig.~\ref{fig:BLSCountsVsF}, we show in Fig.~\ref{fig:LocalVsPropSWs}(a) \& (b) a comparison of the power dependence of the peaks at both overlapping and separated laser spots, again for a field of 600 mT. While again all harmonics can be observed in the overlapping case, the number of peaks is dramatically reduced when the pump and probe spots are separated by 1~$\mu$m, with essentially only the 8 and 9 GHz peaks surviving. We are hence lead to conclude that only at these two frequencies do the laser pulses generate propagating SWs of any magnitude; the magnetodynamics at all lower harmonics are consequently of a local nature and the higher harmonics are more strongly damped out.

In Fig.~\ref{fig:LocalVsPropSWs}(c) \& (d) we plot the peak counts of the 8 GHz mode vs.~fs-laser fluence both at the spot where the fs-laser irradiates the film and 1 $\mu$m away. We have averaged the results from five different measurement spots in each case (both the pump and the probe lasers were moved to different locations). While the BLS counts clearly show a non-linear, square-like, dependence on fluence, the data is not well fitted by a parabola. Instead, as described below, we develop a simple model where we incorporate a Bloch law dependence for how the magnitude of the demagnetization pulse depends on the effective magnon temperature. 

Our theoretical model assumes that each optical pulse rapidly increases the temperature of the magnetic subsystem in linear proportion to the laser fluence. This, in turn, causes a rapid reduction of the magnetization $\delta M_s$, followed by a slower recovery where the magnetization relaxes back to its original value. In the low temperature limit, the temperature dependence of the magnetization should follow a Bloch $T^{3/2}$ law and consequently $\delta M_s = A \left(t_r^{3/2}-(t_r+ F)^{3/2}\right),$
where F is the laser fluence, and $A$ are $t_r$ are coefficients defined by the magnetization $M_s (T=0)$, the Curie and room temperatures, and the heating efficiency of the laser. One can fit BLS counts as $\delta M_s^2 = A^2 \left(t_r^{3/2}-(t_r+ F)^{3/2}\right)^2$.
The model is found to fit both curves exceptionally well. We hence conclude that \emph{i}) heating of the magnon population in direct proportion to the laser power describes the rapid demagnetization very well, and \emph{ii}) BLS counts are a very sensitive probe to study the power dependence of rapid demagnetization. 

\subsection{Spatial mapping of spin wave amplitude:}

\begin{figure}[h]
    \begin{center}
        \includegraphics[width=\linewidth]{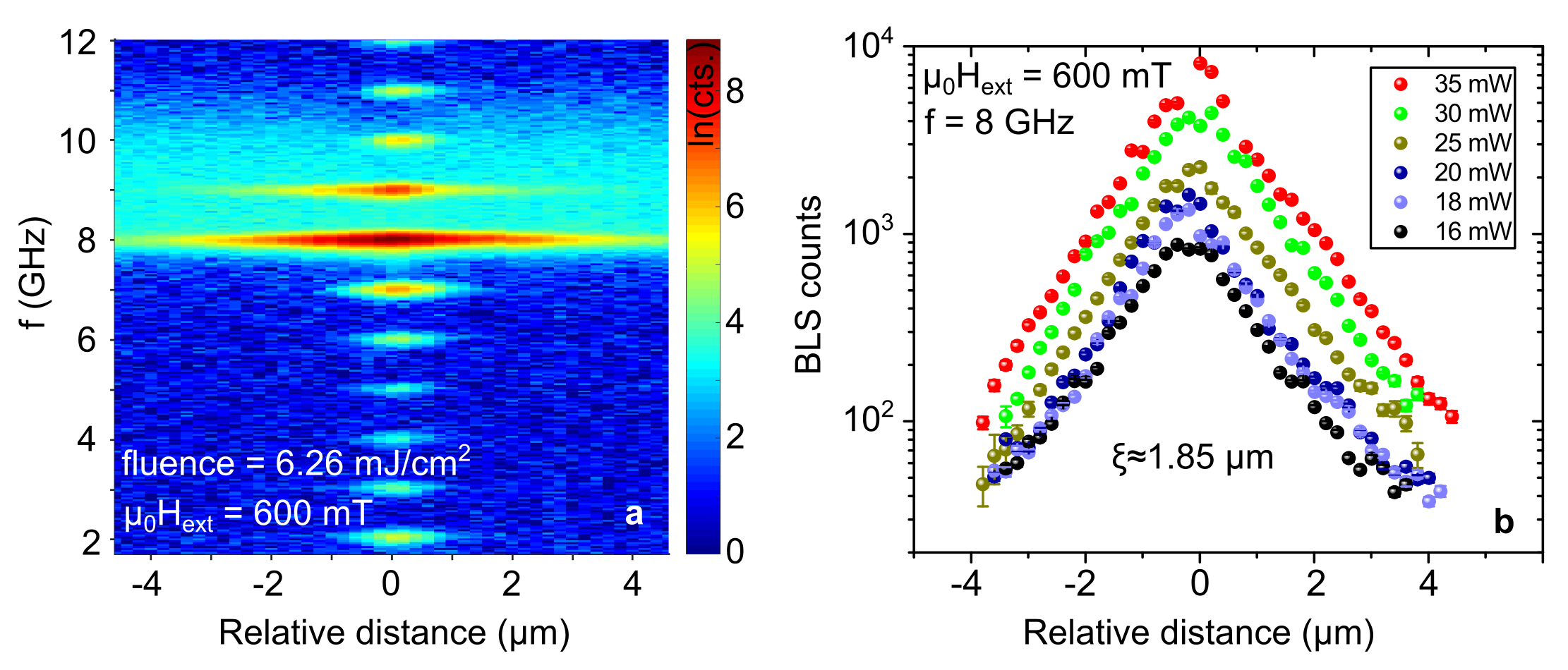}
        \caption{\textbf{Line profiles of the spin wave intensity. }\textbf{a.} BLS counts (log scale) vs.~relative distance from the fs-laser pump spot along the axis of SW propagation. \textbf{b.} BLS counts at 8~GHz as a function of the relative distance between the pump and probe beams for six different power levels. $\xi$ is the spin wave propagation length calculated from an exponential fit to all the data.
        }
        \label{fig:BLSLinescans}
    \end{center}
\end{figure}

\begin{figure}[b]
    \begin{center}
        \includegraphics[width=\linewidth]{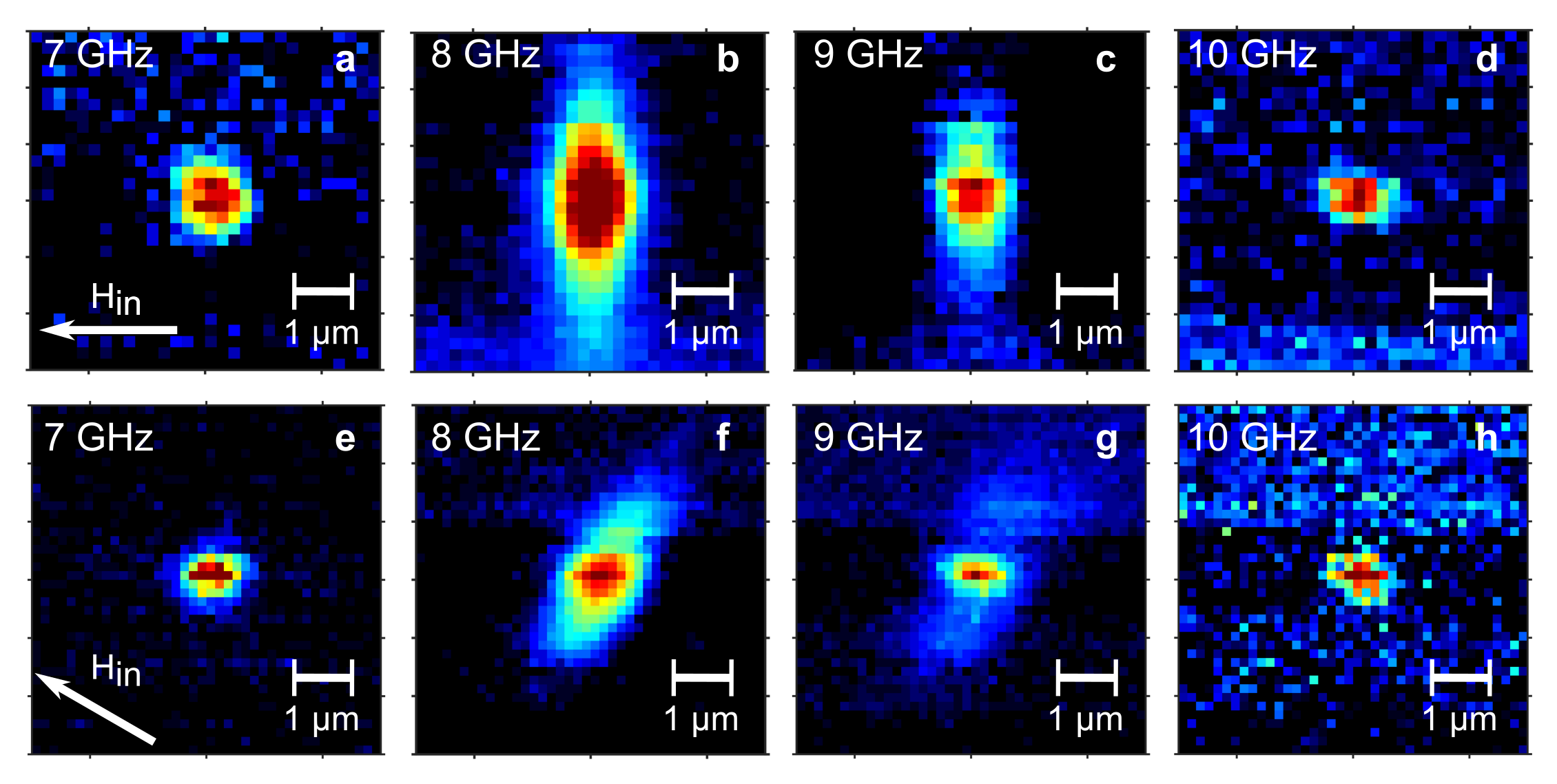}
        \caption{\textbf{2D spatial profiles of the spin wave excitation. } (a)--(d) Area maps of the spin wave intensities at 7, 8, 9 and 10~GHz. (e)--(h) show the same type of maps after the in-plane component of the applied field has been rotated 30 degrees clockwise.}
        \label{fig:2Dmaps}
    \end{center}
\end{figure}

To study in detail the propagation characteristics of all excited SW modes we measured the spatial extension of the spectra at a 600 mT applied field. Fig.\ref{fig:BLSLinescans}~(a) shows a lateral scan of the BLS counts relative to the pump spot along the direction perpendicular to the in-plane component of the applied magnetic field.

At 8 GHz, SWs propagate well beyond 4 $\mu$m on either side of the excitation spot before drowning in the thermal background. Similarly, SWs at 9 GHz propagate up to 3 $\mu$m on either side. In contrast, the modes below 8 GHz are localized to the pump region, and the modes above 9 GHz show no discernible propagation. Fig.\ref{fig:BLSLinescans}~(b) shows the peak counts for 8~GHz (at 600~mT) for different pump powers as a function of the vertical coordinate. 
By fitting the profiles to an exponential decay function of the form $\exp(-2x/\xi)$, where $\xi$ is the SW propagation length, we obtain $\xi = 1.85~\mu$m. It is noteworthy that the SW decay rate, or the damping constant, remains the same for all pump powers.

\begin{figure*}
    \begin{center}
        \includegraphics[width=0.8\textwidth]{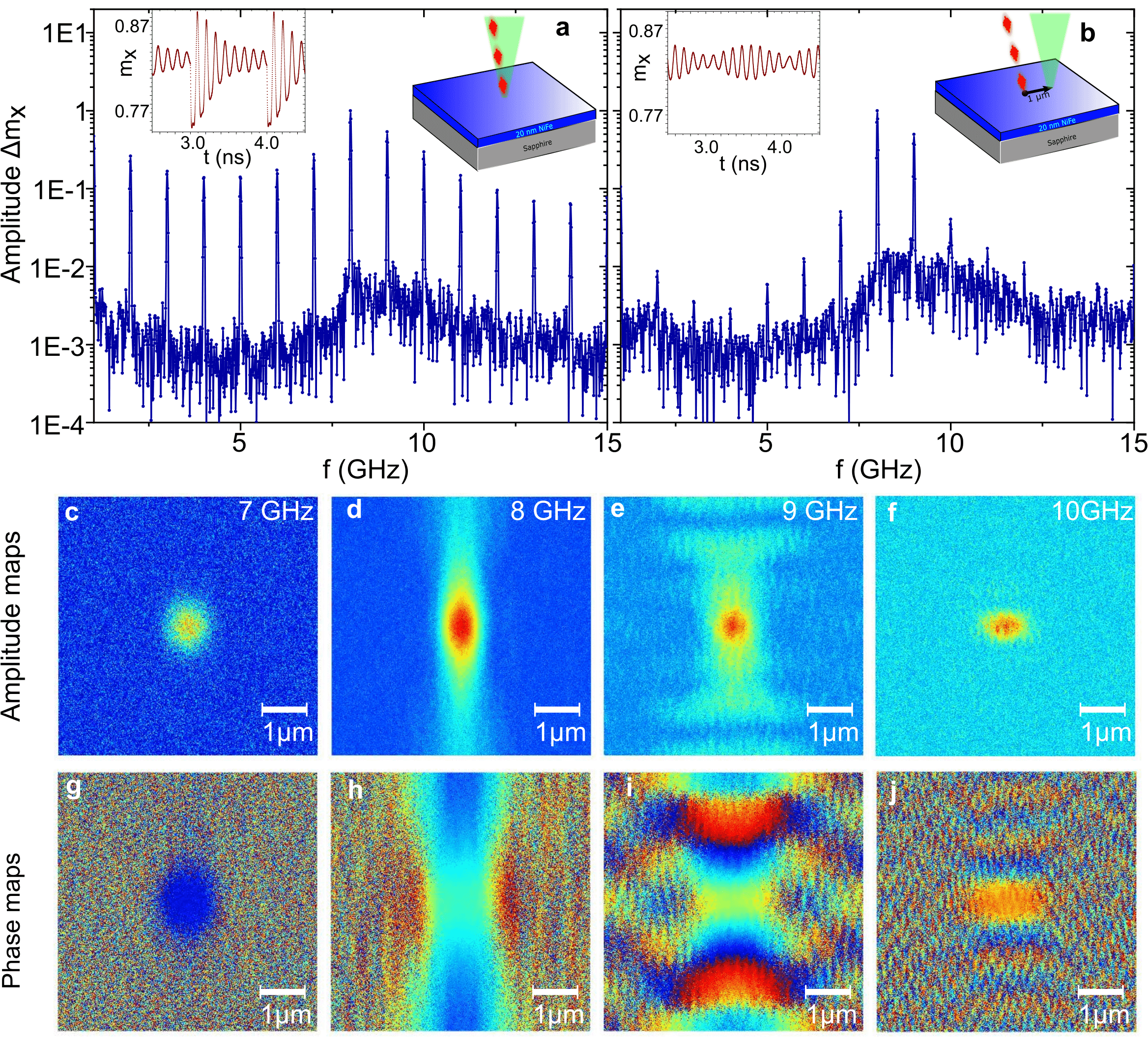}
        \caption{\textbf{Micromagnetic simulations. } Simulated spectral density of the magnetization dynamics: \textbf{a.} at the pump location, and \textbf{b.} at a distance of $1~\mu$m from the pump location. Insets show the instantaneous temporal evolution of the magnetization. The simulated amplitude (\textbf{c, d, e, f}) as well as the phase profiles (\textbf{g, h, i, j}) for the strongest harmonics at 7, 8, 9, and 10 GHz are shown in the lower subfigures. }
        \label{fig:uMag}
    \end{center}
\end{figure*}

Full 2D spatial profiles of the SW modes can then be obtained by raster scanning over the sample. Fig.\ref{fig:2Dmaps}~(a)-(d) are the respective mode profiles of the 7, 8, 9 and 10~GHz peaks. The mode at 7~GHz is clearly localized at the pump spot. At 8 GHz the pump harmonic coincides approximately with the broadband ferromagnetic resonance peak and at 9~GHz with higher wave vector SWs. High amplitude SWs are then excited from the center and propagate in the direction perpendicular to the in-plane component of the applied field. At 10 GHz, we do not find any measurable evidence for SW propagation. As SW frequencies of 10 GHz in a field of 0.6 T correspond to wave vectors just above the cut-off of the resolution of our BLS (see Fig.2a) we cannot say for certain that no SW propagation takes place at these high wave vectors. However, as the wave-vector cut-off for excitation and detection should depend similarly on spot size, and the spot size of the fs-laser ($\sim$ 800 nm) is larger than that of the BLS laser ($\sim$ 350 nm), this is unlikely. Additionally, in Fig.\ref{fig:2Dmaps}~(e)-(h) we demonstrate that it is also possible to steer the propagation direction by rotating the in-plane-component of the applied magnetic field. 

\subsection{Micromagnetic simulations:}

To gain further insight into the nature of the SW emission, we performed micromagnetic simulations at an applied external field value of 0.6 T.
The micromagnetic simulations were performed on a 5.12*5.12 $\mu$m$^2$ permalloy film of 20~nm thickness using Mumax3\cite{vansteenkiste2014aip}. Periodic boundary conditions were used to suppress finite size effects expected for this sample.  A saturation magnetization $M_s$ of 781.75 kA/m extracted experimentally from thermally excited FMR, exchange stiffness \textit{A} of 11.3 pJ/m and damping of 0.01 are the parameters used in the simulations. As in the experiment, an external field was applied at an oblique angle of $82^\circ$.

To mimic the optical-pump effect spot used in the experiment, an instantaneous reduction of the saturation magnetization followed by a slower recovery where the magnetization relaxes back to its original value, applied at clock frequency of 1 ns in form of subtracting or adding the demagnetization tensor corresponding to the magnetization state at both the demagnetized and the recovered states. The pump-beam, as in the experiment, had a Gaussian profile with FWHM of 800 nm. 

Number of magnons per unit volume (density of a magnetic moment carried by spin waves), at a certain frequency $\omega$ was extracted from the results of simulations using Eq.(S19) in \cite{dzyapko2017magnon}:
\begin{equation}
N(\omega) =\frac{M_s}{2 \mu_B} \mathbf{m}^*(\omega)\cdot\left[\mathbf{m}_0 \times \mathbf{m}(\omega)\right],   
\end{equation}
where $\mathbf{m_0}$ is a unit vector, in the direction of the equilibrium magnetization , and $\mathbf{m}(\omega)$ - dynamic magnetization in a Fourier space. Our micro-magnetic simulations show almost perfect quadratic dependence of the magnon population at 8 GHz on induced demagnetization $\delta M$.

The results were found to closely reproduce the overall behaviour of the experimental spectra and the spatial profiles of the experimentally observed SW modes. Fig.\ref{fig:uMag} shows the spectrum of the harmonics excited at the pump spot (a) and at a distance 1 $\mu$m away (b). The agreement with the experimental results is quite remarkable. While the magnetization dynamics at the laser spot is dominated by the FMR response, all other harmonics of 1 GHz are clearly present. At low frequencies, the amplitude decreases with harmonic number towards a shallow minimum at $n=$ 4--5 after which the amplitude again increases towards the FMR response; at frequencies above FMR the amplitude at the laser spot again decreases exponentially with $n$. As in the experiment, the spectral response is dramatically different 1 $\mu$m away from the laser spot, where the FMR peak and the first propagating SW mode dominate entirely and all other peaks are strongly suppressed. 

The insets show time traces of the $x$ component of the magnetization at the two locations. While both time traces show a continuous SW intensity, the dynamics at the laser spot is more strongly affected by each pulse, whereas the response at 1 $\mu$m away varies smoothly and exhibits beating from the sum of the two dominant spectral lines.

Fig.\ref{fig:uMag} c-h shows the spatial maps of the SW amplitudes and phase at modes corresponding to 7, 8, 9 and 10~GHz.
As observed in the experiment, the SWs at 7 GHz correspond to a localized mode, as its frequency falls below the FMR. All higher frequencies correspond to propagating spin waves with wave-vectors perpendicular to the in-plane component of the applied magnetic field. 

\section{Conclusions and Outlook}
In conclusion, we demonstrate, using a combined diffraction-limited BLS  SW detection scheme with a diffraction-limited high repetition rate (1 GHz) fs-laser inducing rapid demagnetization, continuous coherent SW emission over a wide range of fields and frequencies. The highly efficient SW generation only occurs at higher harmonics of the 1 GHz laser repetition rate, where the SWs that are in phase with the incoming laser pulses are coherently amplified,  while all other SWs are left unaffected. The high sensitivity of our BLS microscope resolves spatial maps of both localized and propagating SWs. We demonstrate that a specific SW wave vector could be coherently amplified, and steer there direction of propagation, by tuning the SW dispersion, through the magnitude and direction of the applied magnetic field.

In contrast to TR-MOKE measurements, where the signal increases linearly with laser power, we observe a nonlinear, stronger than square dependence of the  BLS  counts.   The deviations from a square dependence can be fully accounted for by incorporating a  Bloch law dependence of how the step in the demagnetization field relates to the instantaneous magnon temperature. 

Our results clearly demonstrate the versatility and benefits of using BLS  microscopy for the study of rapid demagnetization stimulated excitation of continuous spin waves, and will enable a wide range of additional photo-magnonic studies of magnetic thin films, magnonic and spintronic devices. While our study used a single excitation spot, it will be straightforward to extend to multiple spots and/or shape the SW excitation region using optical techniques such as spatial light modulation. The possibility of adding sustained SW excitation to the studies of Spin-Transfer Torque (STT) and Spin Hall Effect (SHE) driven nano-scale devices is particularly intriguing as SWs play a key roled in their operation. Outside of these immediate applications, using high repetition rate femtosecond lasers for THz radiation emission in ferromagnet/heavy metal bilayers will likely greatly increase their THz emission intensity. 

\section*{Authors contributions}
A.A.A., S.M., A.A. and J.\AA . designed the experiment. S.M. fabricated the samples. A.A. designed and assembled the optics related to the fs-laser, with assistance from D.H., and characterized the optical parameters. S.M. and A.A.A. conducted most of the measurements and analysis. A.A.A. performed magnetic simulations with support from R.K. and M.D. who assisted with their theoretical expertise. All authors contributed to the data analysis and co-wrote the manuscript.
\begin{acknowledgments}
This work was supported by the Knut and Alice Wallenberg Foundation, the Swedish Research Council, and the Horizon 2020 research and innovation programme (ERC Advanced Grant No.~835068 "TOPSPIN").
\end{acknowledgments}
\section*{Data availability}
The data that support the findings of this study are available from the corresponding author upon reasonable request.
\section*{References}
\bibliography{referencesNatureComm}

\end{document}